\documentclass[preprint, aps, nofootinbib]{revtex4}
\pdfoutput=1
\usepackage{graphicx}
\usepackage{epsfig}
\usepackage{amsmath}
\usepackage{amsfonts}
\usepackage{amssymb}
\usepackage{color}%
\usepackage{dcolumn}
\usepackage{slashed}
\newcommand{\be}{\begin{equation}}
\newcommand{\ee}{\end{equation}}
\newcommand{\ba}{\begin{eqnarray}}
\newcommand{\ea}{\end{eqnarray}}

\newcommand{\mt}[1]{\textrm{\tiny #1}}

\newcommand{\uh}{u_\mt{H}}
\newcommand{\rh}{r_\mt{H}}

\newcommand{\cf}{{\cal F}}
\newcommand{\cb}{{\cal B}}
\newcommand{\ch}{{\cal H}}
\newcommand{\cn}{{\cal N}}

\newcommand{\bea}{\begin{eqnarray}}
\newcommand{\eea}{\end{eqnarray}}

\def\ft#1#2{{\textstyle{\frac{\scriptstyle #1}{\scriptstyle #2} } }}

\def\0{{\sst{(0)}}}
\def\1{{\sst{(1)}}}
\def\2{{\sst{(2)}}}
\def\3{{\sst{(3)}}}
\def\4{{\sst{(4)}}}
\def\5{{\sst{(5)}}}
\def\6{{\sst{(6)}}}
\def\7{{\sst{(7)}}}
\def\8{{\sst{(8)}}}
\def\sst#1{{\scriptscriptstyle #1}}

\begin{document}
\title{Thermoelectric conductivities, shear viscosity, and stability  in an anisotropic linear axion model}
\author{Xian-Hui Ge$^{1}$}
\email{gexh@shu.edu.cn}
\author{Yi Ling$^{2}$}
\email{lingy@ihep.ac.cn}
\author{Chao Niu$^{2}$}
\email{niuc@ihep.ac.cn}
\author{Sang-Jin Sin$^{3,4}$}
\email{sjsin@hangyang.ac.kr}
\affiliation{${}^{1}$Department of Physics, Shanghai University, 200444 Shanghai, China\\
${}^{2}$Institute of High Energy Physics,
Chinese Academy of Sciences, Beijing 100049, China\\
${}^{3}$Department of Physics, Hanyang University, Seoul 133-791, Korea\\
${}^{4}$School of Physics, Korea Institute for Advanced Study, Seoul 130-722, Korea
}

\begin{abstract}
We study thermoelectric conductivities and shear viscosities in a holographically anisotropic model,
 which is dual to a spatially anisotropic $\mathcal{N}=4$ super-Yang-Mills theory at finite chemical potential.
Momentum relaxation is realized through perturbing the linear axion field.
 Ac conductivity exhibits  a coherent/incoherent metal transition. Deviations from the Wiedemann-Franz law  are also observed in our model.
The longitudinal shear viscosity for prolate anisotropy
 violates the  bound conjectured by Kovtun-Son-Starinets.
We also  find that thermodynamic  and dynamical instabilities are not always equivalent by examining the Gubser-Mitra conjecture.
\end{abstract}

\maketitle




\section{Introduction }
One of the advantages of holography is that it provides a nonperturbative method for calculating transport coefficients of strongly coupled systems. Transport coefficients of anisotropic plasma are of interest because
the quark-gluon plasma created in the RHIC and LHC is actually anisotropic and nonequilibrium during the period of time $\tau_{out}$ after the collision. The neutral anisotropic black brane solution at zero temperature was found originally in Ref. \cite{liwei}, and at nonzero temperature was constructed from type-IIB supergravity by Mateos and Trancanelli  \cite{mateos,mateos1}. Interestingly, the shear viscosity longitudinal to the direction of anisotropy violates the viscosity bound \cite{rehban},  which is  up to now the first example of such violation in Einstein gravity  \cite{kss}.

In Refs. \cite{cgs,cgs1}, the R-charged version of the anisotropic black brane solution was derived via nonlinear Kaluza-Klein reduction of type-IIB supergravity to five dimensions. The nonlinear Kaluza-Klein reduction of type-IIB supergravity to five dimensions leads to the presence of an Abelian field in the action. The introduction of the $U(1)$ gauge field breaks the SO(6) symmetry and thus leads to the excitations of the Kaluza-Klein modes. In addition, we  also consider the analytic continuation in which the anisotropy parameter is taken to be imaginary, resulting in an oblate anisotropy.

In this paper, we will calculate the optical conductivity (longitudinal to the anisotropy direction), dc thermoelectric conductivities, and shear viscosities in this anisotropic system. An accurate realization of thermoelectric conductivities in real condensed matter systems requires us to include mechanisms for momentum dissipation. Recently, it was suggested to introduce momentum relaxation in holography by exploiting spatially dependent sources for scalar operators or using the massive gravity \cite{vegh,withers,andrade,davison,blake1,cgs2,johanna,kuang,blake2,WJP:2014,donnos1,donnos2,bg,bg1,amoretti,amoretti1,ling1,ling14}. The reduction action for the anisotropic black brane used in this paper, is the Einstein-Maxwell dilaton-axion theory. We will show that momentum relaxation can be realized in our model by explicitly breaking the translational invariance (for works on spontaneously symmetry breaking, see Ref. \cite{ling} ). Remarkably, the resistivity shows its linear
  temperature behavior, signaling the presence of ``strange metals."

  On the condensed matter theory side, there is still a lack of a satisfying explanation of the linear temperature dependence of resistivity at sufficiently
    high temperatures in materials such as organic conductors, heavy fermions, fullerenes, vanadium dioxide, and pnitides. The linear temperature dependence of resistivity at high temperature, a signature of a breakdown of the Boltzman theory, is expected when the quasiparticle mean free path $l$ becomes shorter than the lattice parameter $\tilde{a}$. It is the self-consistency of the Boltzmann theory which requires  that the charge carrier transport should satisfy the Mott-Ioffe-Regel (MIR) limit $\sigma_{\rm MIR}\sim k^2_F e^2/\hbar$ with $l\sim \tilde{a}$ or $k_{F}l\sim 1$ \cite{mott,mott1}. The violation of the Mott-Ioffe-Regel limit in the so-called ``bad metals" has led to an assertion that the standard theory of Fermi liquids cannot be used to  describe such strongly correlated systems both below and above the MIR limit.  Bad metals cannot yield a quasiparticle description, because such quasiparticles would have a free path shorter than their Compton wavelength. In a holographic setup, the optical conductivity of bad metals can  cross the MIR bound, so that it can be both weaker and stronger than the MIR limit.
     Recent studies on strange metals from holography can be found for examples in Refs. \cite{sm1,sm2,sm3,sm4,sm5}.

Another purpose of this paper is to study the relations between the Gubser-Mitra (GM) conjecture and the ``wall of stability" \cite{vegh,davison}. The GM conjecture states that gravitational backgrounds with a translationally invariant horizon yield an unstable tachyonic mode precisely whenever the specific heat of the black brane geometry becomes negative \cite{GM}. In other words,   dynamical instabilities are correlated with the local thermodynamic instability of background spacetimes. A lot of evidence has been found for this conjecture \cite{FGM,buchel,gmref,gmref1,gmref2,gmref3}.

 On the other hand, it was found that for the shear modes of  the hydrodynamics in massive gravity \cite{davison}, the dispersion relation at the zero-momentum limit is given by
 \be
 \omega= - i  \tau^{-1}_{rel}+\cdot\cdot\cdot,
 \ee
 where $\tau_{rel}$ denotes the momentum relaxation time scale and the ellipsis represents the momentum-dependent terms. The linear fluctuations are purely decaying modes when $\tau^{-1}_{rel}>0$, indicating that the black brane is stable under such dynamical perturbations.  This is what we mean by ``wall of stability." The wall of stability requires the momentum relaxation time scale to be $\tau_{rel}\geq 0$; otherwise, the dual field will absorb momentum rather than dissipating it.  It may be interesting to relate the question of the wall of stability with the GM conjecture and ask the following question: does the regime with $\tau_{rel}\geq 0$ exactly correspond to the thermodynamic stability regime of the black brane? We will prove that
for the isotropic and homogenous black brane solution given in massive gravity \cite{vegh,davison} and Einstein-Maxwell linear scalar theory \cite{withers}, the dynamically unstable regime partially overlaps with the
  local thermodynamical unstable region, but does not fully satisfy the  GM conjecture. However, for the anisotropic black brane investigated here,
the physics in $\tau_{rel}\geq 0$ does not necessarily suggest the local thermodynamic stability of the black brane.

It is well known that  the AdS black holes have rich phase structures. For neutral black holes with spherical topology in asymptotically AdS
spacetime, there is the so-called Hawking-Page phase transition due to a competing effect between the scale set by the volume of the spacetime and the scale determined by the temperature.
For Reissner-Nordstr$\mathrm{\ddot{o}}$m-AdS (RN-AdS) black holes with $S^{d-1}$ horizon topology, the phase diagram is analogous to the phase structure of van der Waals's liquid-gas system \cite{em,nc,lu1,lu2,lu3,chp}. For event horizons with topology $R^{d-1}$, the black brane phase structure is usually considered as dominated by the black brane phase for all temperatures without any thermodynamic instabilities.

However, in Refs. \cite{cgs,cgs1} we found that a planar black brane does not necessarily need to be thermodynamically stable by demonstrating that an anisotropic black brane has a branch of solution with negative specific heat. One may naturally connect the thermodynamic instability uncovered in this anisotropic but translationally symmetric system  with the GM conjecture \cite{GM}. Therefore, in this paper, we will first compute the dc and optical conductivities with momentum relaxation and then check the GM conjecture.

 The organization of the contents is as follows: In Sec. II, we briefly review the anisotropic black brane solution and its thermodynamic properties. In Sec. III, we calculate the dc and ac conductivities.
 In Sec. IV, we calculate the transverse and longitudinal shear viscosities for the prolate brane solutions ( and  the oblate case just for  completeness of our discussions).
 By analyzing the causality structure, we show that oblate anisotropy with $a^2<0$ leads to superluminal propagation of signals in the boundary theory. Therefore, it is not a physical
 solution to be explored in the holographic setup.
 We discuss   relations between the Gubser-Mitra conjecture and the wall of stability, and extend our discussion to the massive gravity and Einstein-Maxwell linear scalar theories in Sec. V.  We present our conclusions in the last section.

\section{R-charged anisotropic black brane solution}
The five-dimensional axion-dilaton Maxwell gravity bulk action reduced from type-IIB supergravity is written as \cite{cgs,cgs1}
\be\label{5action}
S=\frac{1}{2\kappa^2}\bigg[\int d^5 x \sqrt{-g} \Big( R +12-\ft12(\partial\phi)^2 - \ft12 e^{2\phi} (\partial \chi)^2- \ft14 F_{\mu\nu}F^{\mu\nu}\Big)-2\int d^4x \sqrt{-\gamma}K \bigg]\,,
\ee
where we have set the AdS radius $L=1$, and $\kappa^2=8\pi G=\frac{4\pi^2}{N^2_c}$. The counterterm takes the form
\be
 S_{ct}=\frac{1}{\kappa^2}\int d^4x \sqrt{\gamma}(3-\frac{1}{8}e^{2\phi}\partial_i\chi\partial^i\chi)-\log v\int d^4x \sqrt{\gamma} \mathcal{A},
 \ee
 where $\mathcal{A}$ is the conformal anomaly in the axion-dilaton gravity system, $v$ is the Fefferman-Graham  coordinate, and $\gamma$ is the induced metric on a $v=v_0$ surface.

\begin{figure}[htbp]
 \begin{minipage}{1\hsize}
\begin{center}
\includegraphics*[scale=0.50]{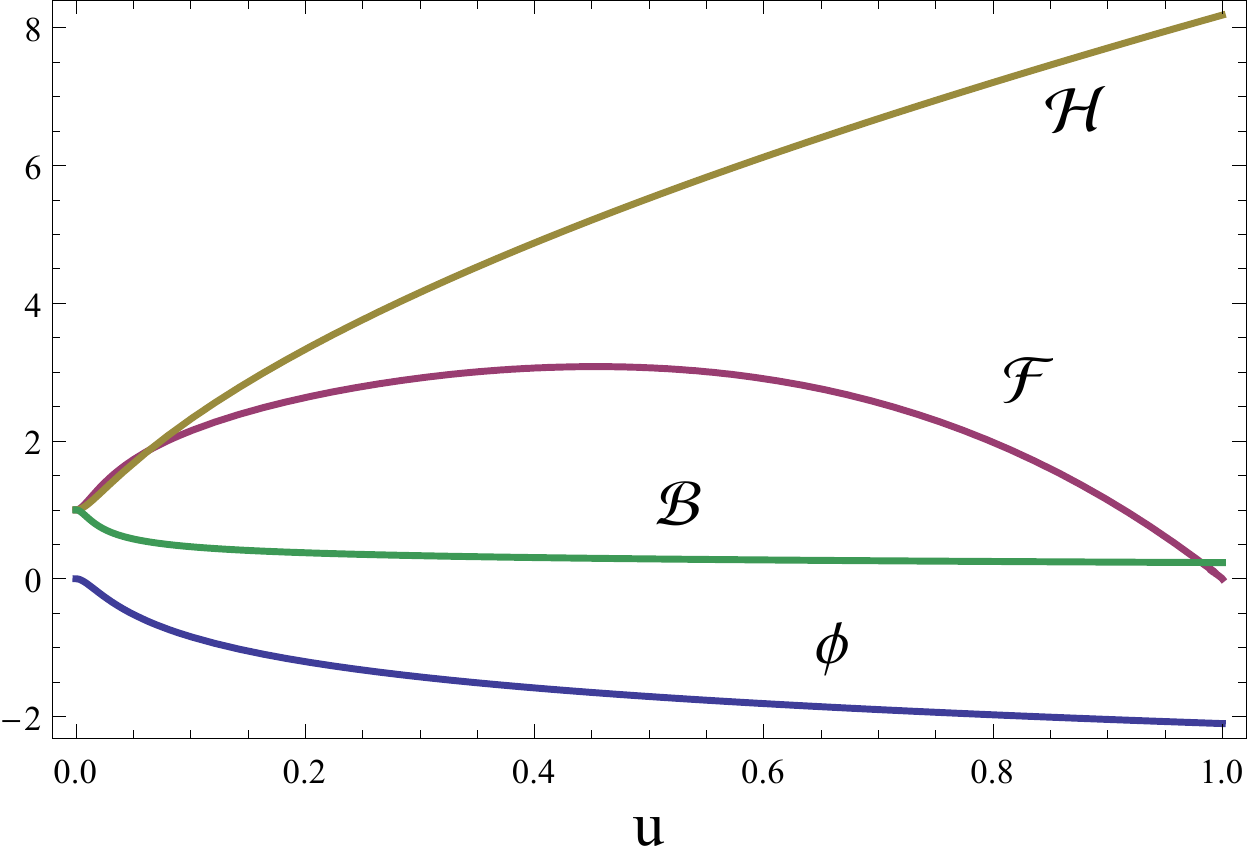}
\includegraphics*[scale=0.50] {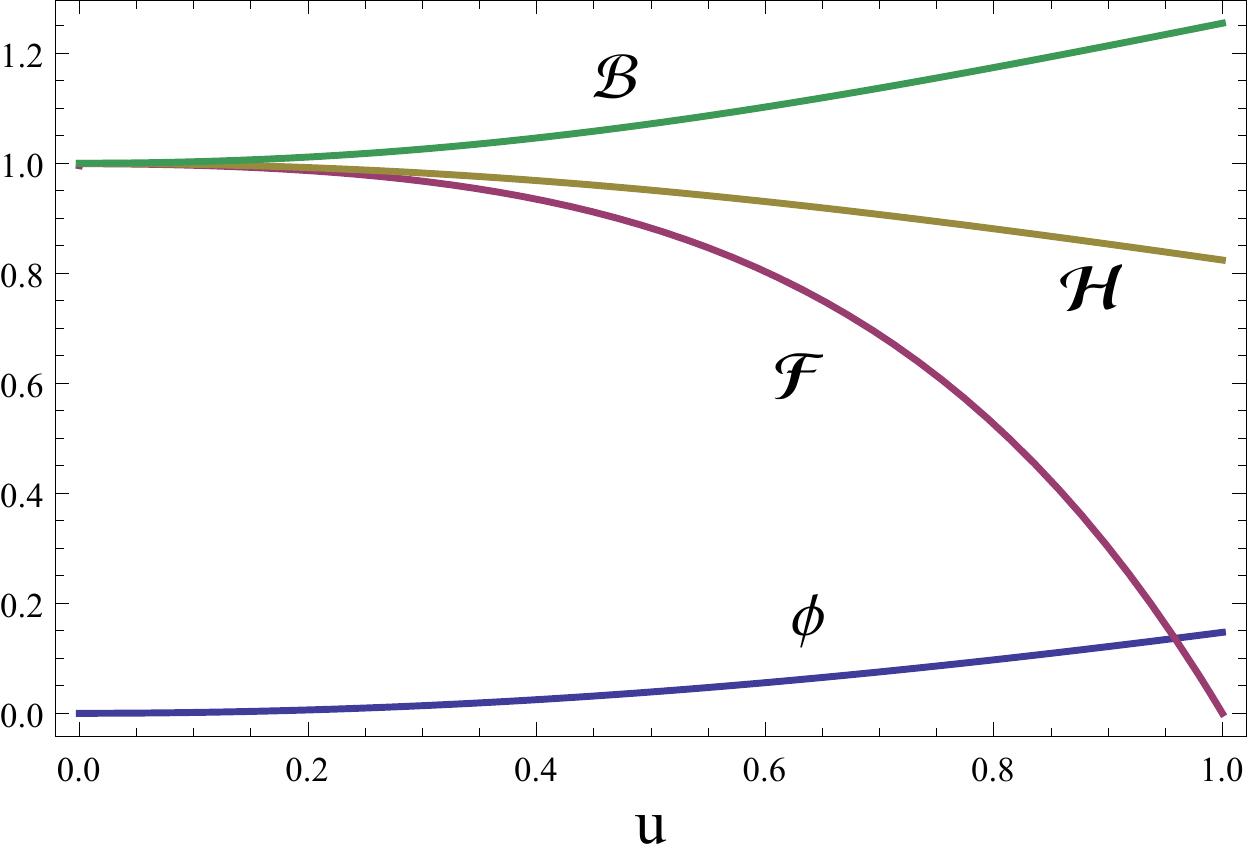}
\end{center}
\caption{ The metric functions for $a=64.06$, $Q=9.76$ (left), which corresponds to the prolate anisotropy; and  $a=1.2i$, $Q=1/10$(right), with $u_H=1$, which depicts the oblate anisotropy.} \label{sol1}
\end{minipage}
\end{figure}

The background solutions with anisotropy along the $z$ direction for the equations of motion are
\begin{eqnarray}\label{metric}
&&ds^2=e^{-\frac{1}{2}\phi}r^2\Big(-\mathcal{F}\mathcal{B}dt^2+dx^2+dy^2+\mathcal{H}dz^2\Big)+\frac{e^{-\frac{1}{2}\phi}dr^2}{r^2\mathcal{F}}\\
&&A=A_t(r) dt,~~~\phi=\phi(r),~~~\chi=a z
\end{eqnarray}
The metric functions $\phi$, $\cf$, $\cb$, and $\ch=e^{-\phi}$ are functions of the radial coordinate $r$ only. The electric potential is given by $A_t(r)=\int^r_{\rh}dr Q\sqrt{\cb}e^{\frac{3}{4}\phi}/r^3$, where $Q$ is an integral constant related to the charge. A dimensionless charge can be introduced by defining $q\equiv \frac{Q}{2\sqrt{3}\rh^3}$, and the physical range of $q^2$ is $0\leq q^2 <2$. The horizon  locates at $r=\rh$ with $\mathcal{F}(\rh)=0$, and the boundary is at $r \rightarrow \infty$ where $\cf=\cb=\ch=1$. The asymptotic $AdS_5$ boundary condition requires the boundary condition $\phi( \infty)=0$.
We note that the above ansatz is invariant under the scaling $t\rightarrow \lambda t$, $x_i \rightarrow \lambda x_i$, $r \rightarrow \lambda^{-1} r$, and $a \rightarrow \lambda^{-1} a$.
The Hawking temperature is given by
\bea
T  &=&   \frac{\rh^2\cf'(\rh) \sqrt{\cb_\mt{H}}}{4\pi} = \sqrt{\cb_\mt{H}}\bigg[\frac{\rh e^{-\frac{\phi_H}{2}}}{16\pi}\bigg(16+\frac{a^2 e^{7\frac{\phi_H}{2}}}{\rh^2}\bigg)-\frac{e^{2\phi_H}q^2 \rh}{2\pi}\bigg] ,
\label{temperature} \,
\eea through the Euclidean method. The entropy density is given by $s=\frac{N^2_c e^{-\frac{5\phi_H}{4}}}{2\pi}\rh^3$. Note that the prolate anisotropy $a^2>0$ corresponds to the metric function $\ch (\rh)>1$, since the metric has a $z$ axis longer than the $x$ and $y$ axes. From an analytical continuation in which the anisotropy parameter $a$ is taken to be imaginary, one can obtain an oblate metric.
 The numerical and semianalytic black brane solution was given in Refs. \cite{cgs,cgs1}.  For example, we  can plot the numerical solutions (\ref{metric}) in Fig. \ref{sol1} for prolate and oblate anisotropy. We will prove in Sec. IV that the oblate anisotropy is unphysical in the holographic setup.

The Ward identity obeys
\be\label{ward}
\nabla^i \langle T_{ij}\rangle=\langle \mathcal{O}_{\phi} \rangle \nabla_j \phi^{(0)}+\langle \mathcal{O}_{\chi} \rangle \nabla_j \chi^{(0)}+F^{(0)}_{ij} \langle J_i \rangle.
\ee
If we consider the background fields only, the translational symmetry is unbroken because $\nabla^i \langle T_{ij}\rangle=0$, with the facts that $\nabla_j \phi^{(0)}=0$, $\langle \mathcal{O}_{\chi} \rangle=0$, and $F^{(0)}_{ij}=0$.  However, the black brane solution is not translationally invariant in the $z$ direction.
 By considering the fluctuations around the background as we will do in later sections, we can prove that $\langle\mathcal{O}_{\chi}\rangle$ is finite. Thus, the Ward identity suggests an approach to holographic momentum relaxation and finite dc response through the spatially dependent source term for the axion.

Actually, the anisotropic black brane has very special thermodynamic  properties, as pointed out  in Refs. \cite{cgs,cgs1}. Considered the prolate anisotropy with $a^2>0$, the black brane suffers thermodynamic instabilities. This can be easily seen from (\ref{temperature}) in the small-horizon-radius limit $\rh \ll 1$; that is to say
\be
T \sim \frac{\sqrt{\cb_\mt{H}}a^2 e^{3\phi_H}}{16 \pi \rh},
\ee
which in turn results in negative specific heat, since $\partial T/\partial \rh <0$. On the other hand, for the larger horizon radii with $\rh \gg 1$, we have
\be
T \sim   \frac{\sqrt{\cb_\mt{H}}a^2 e^{-\phi_H/2}}{16 \pi }\bigg(16-\frac{q^2 e^{2\phi_H}}{2\pi}\bigg)\rh.
\ee
Thus for the prolate anisotropy case, for a fixed temperature there are two branches of allowed black brane solutions: a branch with larger horizon radii and one with smaller. The smaller branch   solution is unstable with negative specific heat. This situation is very similar to the case of Schwarzschild-AdS black holes with a spherical  horizon. In general, the larger black brane branch will appear and will match to the small black brane solution at some temperature $T_{\rm min}$, below which  there is only the thermal gas solution. At some higher temperature, a first-order phase transition  (from thermal gas) to the large black brane branch will take place. So the system will be a doped Mott-like insulator  up to a critical temperature, and then there will be a first-order phase transition to a conducting phase  \cite{rene,bs,bs1}.
In the following, we mainly focus on the prolate anisotropy  case with $a^2>0$.

 \section{DC and optical conductivity with momentum relaxation}
   In what follows, we will first compute the dc conductivity with momentum dissipation and discuss its physical meaning.
In the $r=1/u$ coordinate, the metric can be recast as
\be
ds^2=-g_{tt}(r) dt^2+g_{rr}(r) dr^2+g_{xx}(r) dx^2+g_{yy}(r) dx^2+g_{zz}(r) dz^2,
\ee
where $g_{xx}=g_{yy}\neq g_{zz}(r)$.

It was argued in several papers that  axions having a spatially dependent source leads to the fact that momentum is dissipated  at the linearized level \cite{withers,donos,kim}.
For the scalar type  of metric perturbations,  the independent variables are $h_{tt}$, $h_{tz}$, $h_{xx}=h_{yy}$, $h_{zz}$, $\delta \phi$, $\delta\chi$ together with the $t$ and $z$ components of the gauge
field $A_{\mu}$.  In the zero-momentum limit, it is easy to check that $h_{tz}$, $A_z$, and $\delta \chi$ decouple from other variables.
Therefore, to compute the conductivity with momentum dissipation,
we only need to consider linearized fluctuations of the form
\ba
\delta g_{tz(0)}=h_{tz}(t,r) ,~~~ \delta A_{z}(t,r)=a_z(t,r) ,~~~ \delta \chi=a^{-1}\bar{\chi}(t,r),
\ea
and all the other metric and gauge perturbations vanished.  We observe that the axion only sources momentum relaxation along the anisotropic direction. If we instead consider the metric perturbations $h_{tx}$ and $\delta A_x (t,r)$, we expect that it is metallic along the isotropic direction without any momentum dissipation.     Here we choose the gauge $h_{rz}=0$ and  the electromagnetic perturbation along the $z$ direction at zero momentum. We shall work in the Fourier decomposition
\ba
&& h_{tz}(t,r)=\int \frac{d \omega}{2\pi}e^{-i\omega t} h_{tz}(\omega,r),\\
&& a_z (t,r)=\int \frac{d \omega}{2\pi}e^{-i\omega t} a_{z}(\omega,r),\\
&& \bar{\chi}(t,r)=\int \frac{d \omega}{2\pi}e^{-i\omega t} \bar{\chi}(\omega,r).\ea
The linearized equations of motion corresponding to the $(r,z)$ component of Einstein's equation, the Maxwell equation, and the dilaton equation are given by
\bea
0&=&A'_t a_z+h'_{tz}-\frac{g'_{zz}}{g_{zz}}h_{tz}-\frac{g_{tt}e^{2\phi}\bar{\chi}'}{i\omega},\label{con}\\
0&=& a''_z+\bigg(\frac{g'_{tt}}{2g_{tt}}-\frac{g'_{zz}}{2g_{zz}}+\frac{g'_{xx}}{g_{xx}}-\frac{g'_{rr}}{2g_{rr}}\bigg)a'_z+\frac{A'_t}{g_{tt}}(h'_{tz}-\frac{g'_{zz}}{g_{zz}}h_{tz})+\frac{\omega^2 g_{rr}}{g_{tt}}a_z,\label{az}\\
0&=& \bar{\chi}''+\bigg(\frac{g'_{zz}}{2g_{zz}}+\frac{g'_{tt}}{2g_{tt}}+\frac{g'_{xx}}{g_{xx}}-\frac{g'_{rr}}{2g_{rr}}+2\phi'\bigg)\bar{\chi}'+\frac{\omega^2 g_{rr}}{g_{tt}}\bar{\chi}-\frac{i\omega a^2 g_{rr}}{g_{zz}g_{tt}}h_{tz}, \label{chi}\\
0&=& h''_{tz}-\bigg(\frac{g'_{tt}}{2 g_{tt}}+\frac{g'_{rr}}{2 g_{rr}}+\frac{g'_{zz}}{2 g_{zz}}-\frac{g'_{xx}}{ g_{xx}}\bigg)h'_{tz}+\bigg(\frac{g'_{rr}g'_{zz}}{2 g_{rr}g_{zz}}+\frac{g'_{tt}g'_{zz}}{2 g_{tt}g_{zz}}-\frac{g''_{zz}}{g_{zz}}+\frac{g'^2_{zz}}{2 g^2_{zz}}\nonumber\\[0.7mm]&&-\frac{g'_{xx}g'_{zz}}{ g_{xx}g_{zz}}-\frac{g_{rr}e^{2\phi}a^2}{g_{zz}}\bigg)h_{tz}+A'_t a'_z-i\omega g_{rr}e^{2\phi}\bar{\chi},\label{htztr}
\eea
where the prime denotes a derivative with respect to $r$.
For the sake of convenience, we can eliminate $h_{tz}$ by taking a radial derivative of (\ref{chi}) and substituting the expression for $h'_{tz}$ in  (\ref{az}) and (\ref{chi}). We can introduce a new variable
\be
\tilde{\chi}=\omega^{-1}g_{xx}g^{\frac{1}{2}}_{zz}g^{\frac{1}{2}}_{tt}g^{-\frac{1}{2}}_{rr}e^{2\phi}\bar{\chi}',
\ee
and recast the equation of motion as
\bea
0&=&g^{-1}_{xx}g^{\frac{1}{2}}_{zz}\bigg(g_{xx}g^{-\frac{1}{2}}_{zz}\sqrt{\frac{g_{tt}}{g_{rr}}} a'_z\bigg)'+\omega^2\sqrt{\frac{g_{tt}}{g_{rr}}} a_z-\frac{A'_t Q}{g_{xx}\sqrt{g_{zz}}}a_z-\frac{iQ\sqrt{g_{rr}g_{tt}}}{g^2_{xx}g_{zz}}\tilde{\chi},\label{eq1}~~~\\
0&=&e^{2\phi}g_{xx}g^{\frac{1}{2}}_{zz}\bigg(e^{-2\phi}g^{-1}_{xx}g^{-\frac{1}{2}}_{zz}\sqrt{\frac{g_{tt}}{g_{rr}}} \tilde{\chi}'\bigg)'
+\omega^2\sqrt{\frac{g_{tt}}{g_{rr}}} \tilde{\chi}+a^2e^{2\phi}\bigg(\frac{i g_{xx} A'_t }{\sqrt{g_{zz}}}a_z-\frac{ \sqrt{g_{rr}g_{tt}}}{g_{zz}}\tilde{\chi}\bigg).\label{eq2}\nonumber\\
\eea
Following Refs. \cite{blake1,withers}, we can rewrite the fluctuation equations (\ref{eq1}) and (\ref{eq2}) in the form
\bea
\left(\begin{array}{cc}L_1 & 0 \\ 0 & L_2\end{array}\right) \left(\begin{array}{c}a_z \\ \tilde{\chi}\end{array}\right) +\omega^2\sqrt{\frac{g_{tt}}{g_{rr}}} \left(\begin{array}{c}a_z \\ \tilde{\chi}\end{array}\right)=\mathcal{M}\left(\begin{array}{c}a_z \\ \tilde{\chi}\end{array}\right),
\eea
where $L_1$ and $L_2$ are linear differential operators and $\mathcal{M}$ is the mass matrix,
\ba
\mathcal{M}=\left( \begin{array}{cc} {Q^2 \sqrt{g_{rr}g_{tt}}}/(g^2_{xx} g_{zz}) & {i Q \sqrt{g_{rr}g_{tt}}}/(g^2_{xx} g_{zz}) \\ -{i a^2 Q e^{2\phi}\sqrt{g_{rr}g_{tt}}}/g_{zz} & {a^2 e^{2\phi}\sqrt{g_{rr}g_{tt}}}/{g_{zz}}\end{array}\right).
\ea
Clearly, there exists a massless mode, because $\det\mathcal{M}=0 $. Let us introduce the following linear combinations:
\bea
\lambda_1&=&b^{-1}(r)\left(e^{2\phi} a_z +\frac{Q}{i a^2 g^2_{xx}} \tilde{\chi}\right),\\
\lambda_2&=&b^{-1}(r)\left(\frac{Q^2}{a^2 g^2_{xx}} a_z -\frac{Q}{i a^2 g^2_{xx}}\tilde{\chi} \right),
\eea
where
\ba
b(r)=e^{2\phi} +\frac{Q^2}{ a^2 g^2_{xx}}.
\ea
Then, we obtain the master equation for the massless mode $\lambda_1$
\ba
\bigg(e^{-2\phi}g_{xx}\sqrt{\frac{g_{tt}}{g_{rr}g_{zz}}}b(r) \lambda'_1-\sqrt{\frac{g_{tt}}{g_{rr}g_{zz}}}c(r) \lambda_2\bigg)'+\omega^2 b(r)e^{-2\phi}g_{xx}\sqrt{\frac{g_{rr}}{g_{tt}g_{zz}}}\lambda_1=0,
\ea
where $c(r)=(g^2_{xx}e^{2\phi})'/(e^{2\phi}g_{xx})$. We can easily find that the following quantity is radially conserved at zero frequency:
\be \label{pi}
\Pi=e^{-2\phi}g_{xx}\sqrt{\frac{g_{tt}}{g_{rr}g_{zz}}}b(r) \lambda'_1-\sqrt{\frac{g_{tt}}{g_{rr}g_{zz}}}c(r) \lambda_2.
\ee
The dc membrane conductivity to each radial slice is defined as
\be\label{dc}
\sigma_{DC}(r)=\lim_{\omega\rightarrow 0}-\frac{\Pi}{i\omega \lambda_1}\bigg|_{r}.
\ee
It was proved in Refs. \cite{blake1,withers} that $\sigma_{DC}(r)$ does not evolve radially [i.e.$\sigma_{DC}(\infty)=\sigma_{DC}(r_H)$].  So it can be evaluated at the horizon. In order to evaluate (\ref{dc}), we note that the ingoing boundary
conditions for the fields are given by
\bea\label{ingoing}
a_z&=&(r-\rh)^{-i\omega/(\cf'(\rh)\sqrt{\cb(\rh)})}[a^H_z+\mathcal{O}(r-\rh)],\\
\tilde{\chi}&=&(r-\rh)^{-i\omega/(\cf'(\rh)\sqrt{\cb(\rh)})}[\tilde{\chi}^H_z+\mathcal{O}(r-\rh)].
\eea
Substituting the metric functions (\ref{metric}) and (\ref{ingoing}) into (\ref{dc}), we finally obtain
\be\label{DCsig}
\sigma_{DC}=\rh e^{\frac{\phi(\rh)}{4}}\left(1+12 \rh^2 e^{-\phi(\rh)}\frac{q^2}{a^2}\right),
\ee
where $q=\frac{Q}{2\sqrt{3}\rh^3}$. The first term in the round brackets is the conductivity due to the pair production in the dual field theory,
The dc conductivity can also be recast as
\bea \label{sigdc}
\sigma_{DC}&=&(g_{xx}g_{yy}g^{zz})^{1/2}\bigg|_{r=\rh}+\frac{q^2}{a^2 e^{2\phi}(g_{xx}g_{yy}g_{zz})^{1/2}}\bigg|_{r=\rh} \nonumber\\
&=&\rh e^{\frac{\phi(\rh)}{4}}+\frac{Q^2}{a^2\rh^3}e^{-3\phi(\rh)/4}.
\eea
This result is consistent with Refs. \cite{bg,donos}. We will provide an alternative calculation on the dc conductivity later as a consistent check.
 The dc conductivity can be related to a scattering time $\tau_{rel}$ by \cite{amoretti}
 \be
 \sigma_{DC}=(g_{xx}g_{yy}g^{zz})^{1/2}\bigg|_{r=\rh}+\frac{Q^2}{\epsilon+P_{z}}\tau_{rel}.
 \ee
The scattering rate is given by
\be
\Gamma=\tau^{-1}_{rel}=\frac{s a^2}{4\pi(\mathcal{E}+P_{z})}.
\ee
For the prolate solution, we have $\tau_{rel}>0$. For the case  $\tau_{rel}<0$,  the wall of stability will be violated.  As stated in the Introduction, the scattering rate is the characteristic time scale of
momentum relaxation in the dual field theory. When $\tau_{rel}<0$, the metric fluctuations ( quasinormal modes) absorb momentum,  resulting in the fact that small perturbations of the state will grow exponentially in time.\footnote{ In the dual gravitational picture, the unstable quasi-normal modes are identified with unstable uniform plasma with respect to certain nonuniform perturbation \cite{gb1,gb7}. On the gravitational side, this instability seems similar in certain respects to the Gregory-Laflamme instability for black strings\cite{gb7}.}  Therefore, we demand that $\tau_{rel} \geq 0$ for stability. 
 \begin{figure}[htbp]
 \begin{minipage}{1\hsize}
\begin{center}
\includegraphics*[scale=0.4] {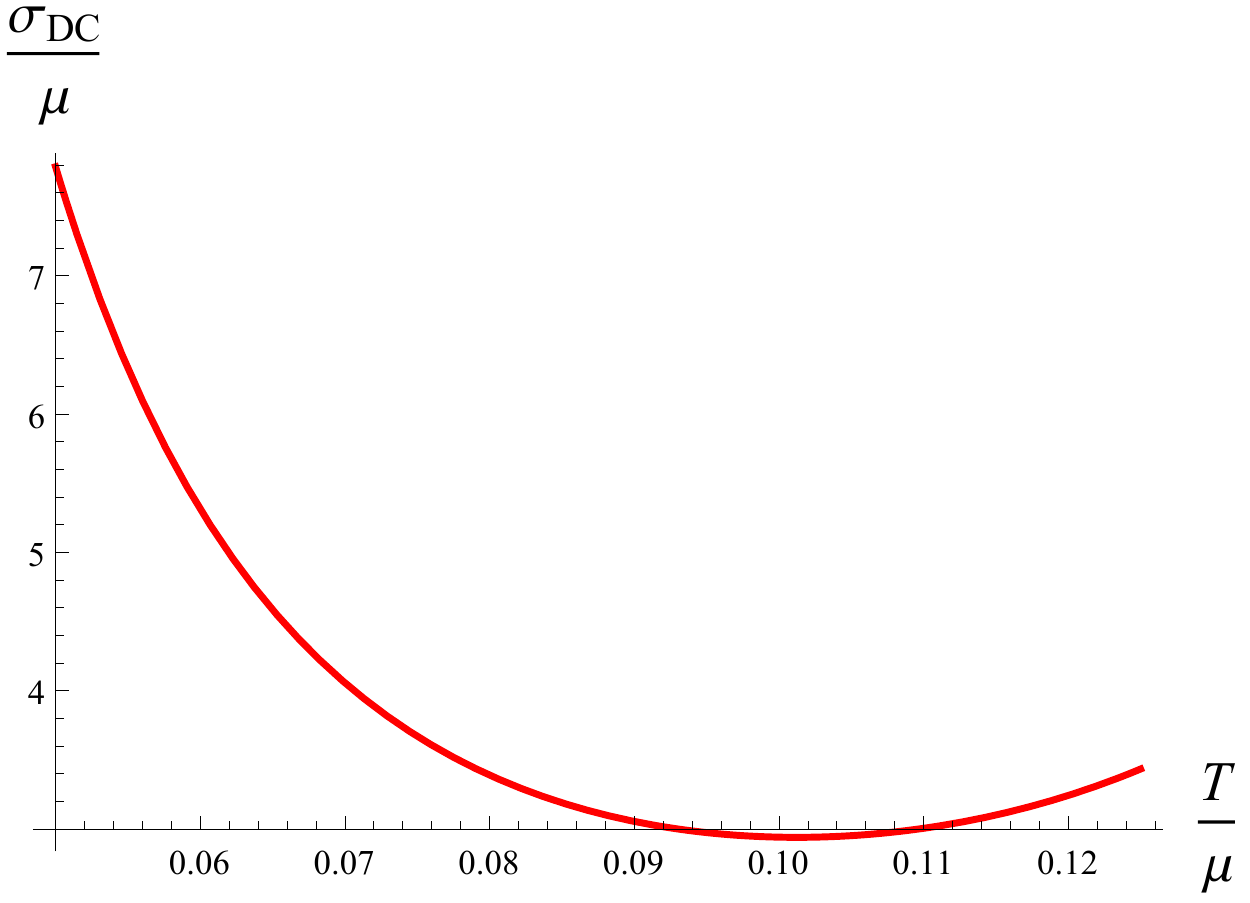}
\includegraphics*[scale=0.4] {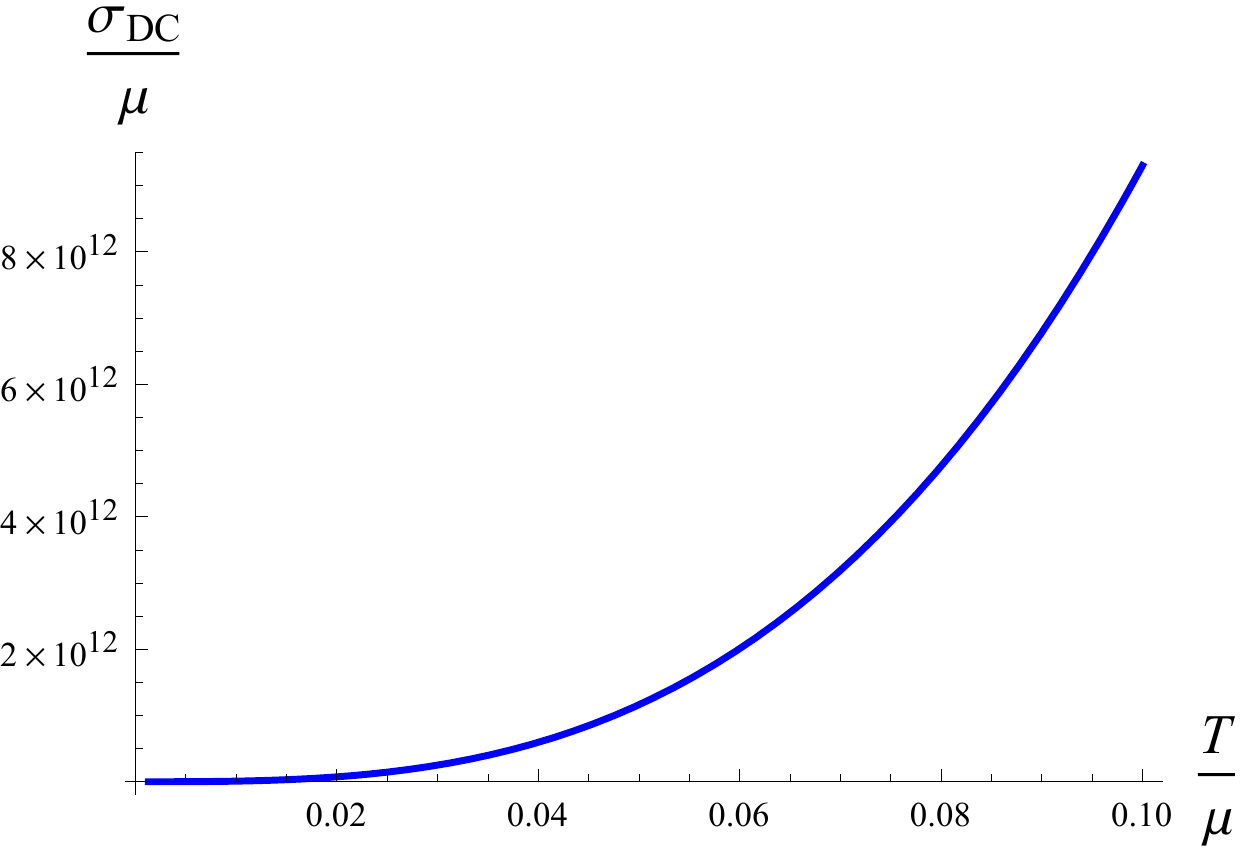}
\includegraphics*[scale=0.4] {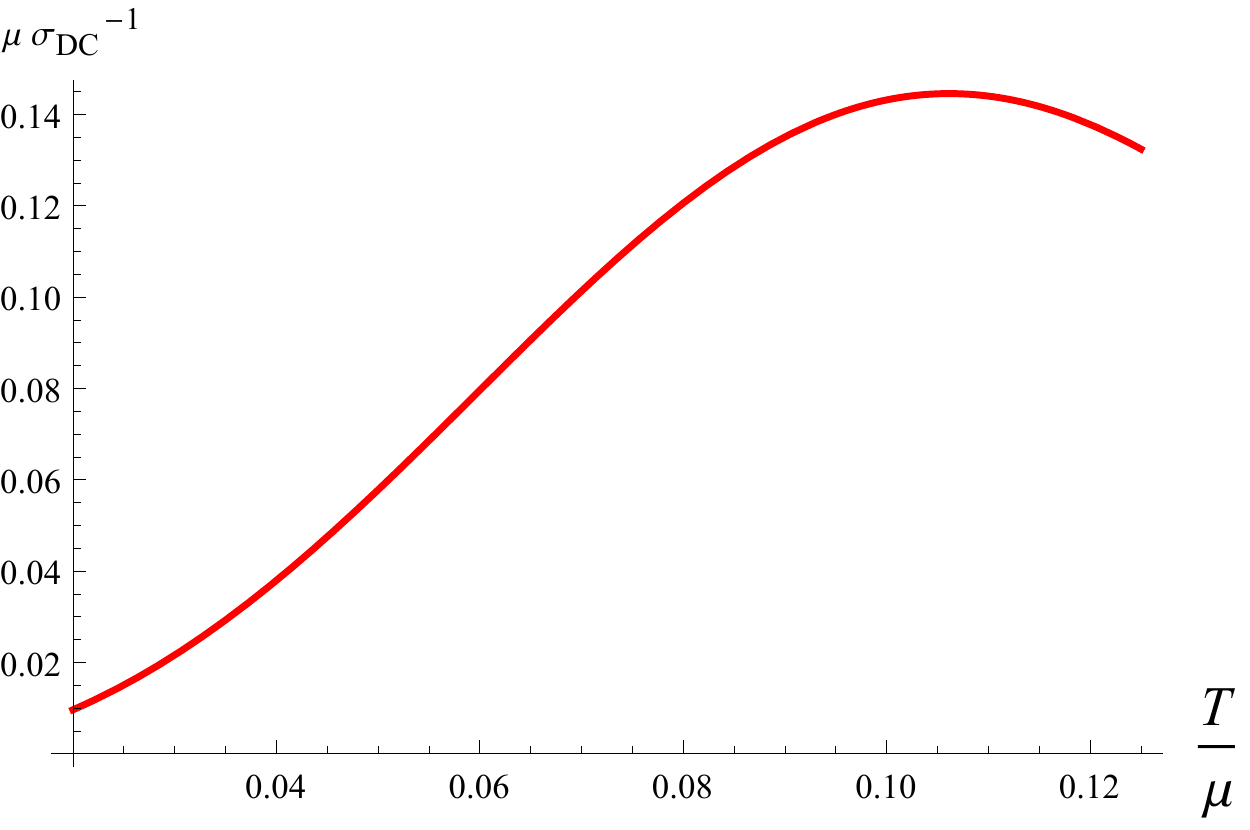}
\end{center}
\caption{ Left: DC conductivity as a function of the  temperature corresponding to the smaller-horizon-radius branch, where we set $q=1.414$ and $a=0.1$.  Middle: DC conductivity as a function of the  temperature in the thermodynamically stable regime, where we set $q=1.414$ and $a=0.1$. Right: The resistivity $\rho\sim \rho_{\rm MIR} T$ shows its linear temperature dependence in the small-horizon regime and that it can violate the Mott-Ioffe-Regel limit at high temperatures. } \label{Tsigma}
\end{minipage}
\end{figure}

To the linear order, the Ward identity (\ref{ward}) reduces to
\bea
 \partial_t \delta \langle T_{zz} \rangle=-a \delta \langle \mathcal{O}_{\chi} \rangle+\rho \partial_t a_z.
\eea
We have proved that the linearized contribution to the vacuum expectation value $\langle\mathcal{O}_{\chi}\rangle$ is finite, although at zeroth order $\langle\mathcal{O}_{\chi}\rangle$ is vanishing.

 $\bullet$ {\textit{Dc conductivity with  prolate anisotropy.}}  We would like to  comment on the  temperature dependence of the dc electric conductivity.
It was proven in Ref. \cite{mateos} that the IR geometry of the prolate black brane is asymptotic Lifshitz at zero temperature, and thus the dc electric conductivity obtained here behaves quiet differently from  those with near-horizon geometry $AdS_2\times R^3$ \cite{withers}.
There are two branches in the high-temperature phase: small and large black hole radius.
As can be seen in Fig. \ref{Tsigma} (middle), the conductivity goes up as the temperature increases, which  corresponds to the larger-horizon-radius branch solution of the black brane.  The system undergoes a first-order Hawking-Page-like confinement/deconfinement phase transition as the temperature varies. It has been proved that the smaller-horizon-radius branch of the solution has a negative specific heat, so it is not physically realized.
  If the small-horizon-radius $\rh$ were stable,
we could identify it as the dual to doped ``bad metals," since
its resistivity is linear in temperature as shown in Fig. \ref{Tsigma} (right).
    This is because in the holographic set up, quasiparticle descriptions lose their validity and  dc resistivity can be both weaker and stronger than the MIR limit $\rho_{\rm MIR}\sim \hbar/k^2_F e^2$.
It suggests that we need to stabilize the small-radius branch by twisting
our present model, which suggests a direction to future model building.

In Refs. \cite{rene,bs,bs1}, the authors argued that the confinement phase is dual to the Mott insulators. The  strange metal behaviors can be  obtained by doping a Mott insulator. In our setup, the linear axion field's parameter $a$ might play the role of dopants.

\subsection{Optical conductivity}
In this section, we try to solve Eqs. (\ref{con})-(\ref{htztr}) numerically.
In order to calculate ac   thermo-electric conductivities numerically, we need to evaluate the on-shell action that gives a finite and quadratic function of the boundary values. In general, for the variation of the action, we have
\be
\delta S=\int \partial_{\mu}\bigg(\frac{\partial L}{\partial \partial_{\mu}\varphi^i}\delta\varphi^i\bigg)dr+\int E.O.M. \delta\varphi^i dr.
\ee
We obtain the on-shell action in the momentum space,
\bea
S= \lim_{r\rightarrow \infty}\frac{V_3}{2} \int \frac{d \omega}{2\pi}\sqrt{-g}\bigg[\frac{g_{zz}}{g_{rr}g_{tt}}h'_{tz}h_{tz}-\frac{a'_z a_z}{g_{rr}g_{zz}}-\frac{e^{2\phi}\bar{\chi}'\bar{\chi}}{g_{rr}}+\frac{A'_t a_z}{g_{rr}g_{tt}}h_{tz}- \mathcal{E} h_{tz}h_{tz}\bigg],
\eea
where $\mathcal{E}$ denotes the energy density.  The Green function is defined via $G_{i;j}=\delta^2 S/\delta\varphi^i\delta\varphi^j$. Near the boundary ($r\rightarrow \infty$), the asymptotic solutions go as
\bea
&&h_{tz}=r^2 h^{(0)}_{tz}+ h^{(2)}_{tz}+\frac{1}{r} h^{(3)}_{tz}+\frac{1}{r^2} h^{(4)}_{tz}\cdots,\\
&& a_{z}= a^{(0)}_{z}+\frac{1}{r^2}a^{(2)}_{z}+\cdots,\\
&& \bar{\chi}= \bar{\chi}^{(0)}+\frac{1}{r^2}\bar{\chi}^{(2)}+\frac{1}{r^3}\bar{\chi}^{(3)}+\cdots.
\eea
Note that $a^{(0)}_{z}$ is the source of the electric current $J_z$, and $h^{(0)}_{tz}$ is dual to the source of the energy-momentum tensor $T_{tz}$. In order to compute the ac electric conductivity numerically, we need to impose the ingoing boundary condition at the horizon and adopt the numerical method developed in Refs. \cite{donnos2} and \cite{kim}.
\begin{table}[ht]
\begin{center}
\begin{tabular}{|c|c|c|c|c|c|c|c|c|}
         \hline
~$a/\mu$~ &~$0.1$~&~$0.2$~&~$0.3$~&~$0.4$~&~$0.5$~&~$0.7$~&~$0.8$~&~$1.0$~
          \\
        \hline
~$K/\mu^2$~ & ~$0.751$~ & ~$0.762$~ &
~$0.743$~&~$0.710$~&~$0.698$~& ~$0.634$~& ~$0.628$~& ~$0.614$~
          \\
        \hline
~$\tau\mu$~ & ~$536.715$~ & ~$134.337$~ &
~$64.016$~&~$39.909$~&~$27.456$~& ~$17.966$~& ~$14.924$~& ~$11.385$~
          \\
        \hline
\end{tabular}
\caption{\label{Tablev3}Drude parameters for different $a/\mu$ at
$T/\mu=0.2658$.}
\end{center}
\end{table}
\begin{figure}
\center{
\includegraphics[scale=0.45]{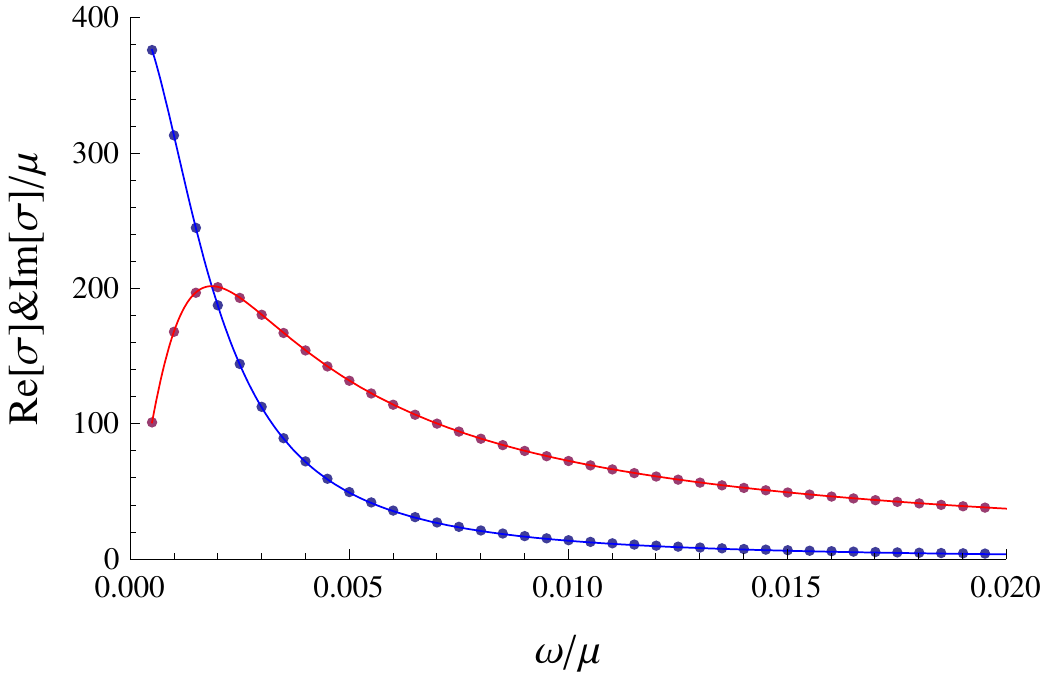}\hspace{0.5cm}
\includegraphics[scale=0.45]{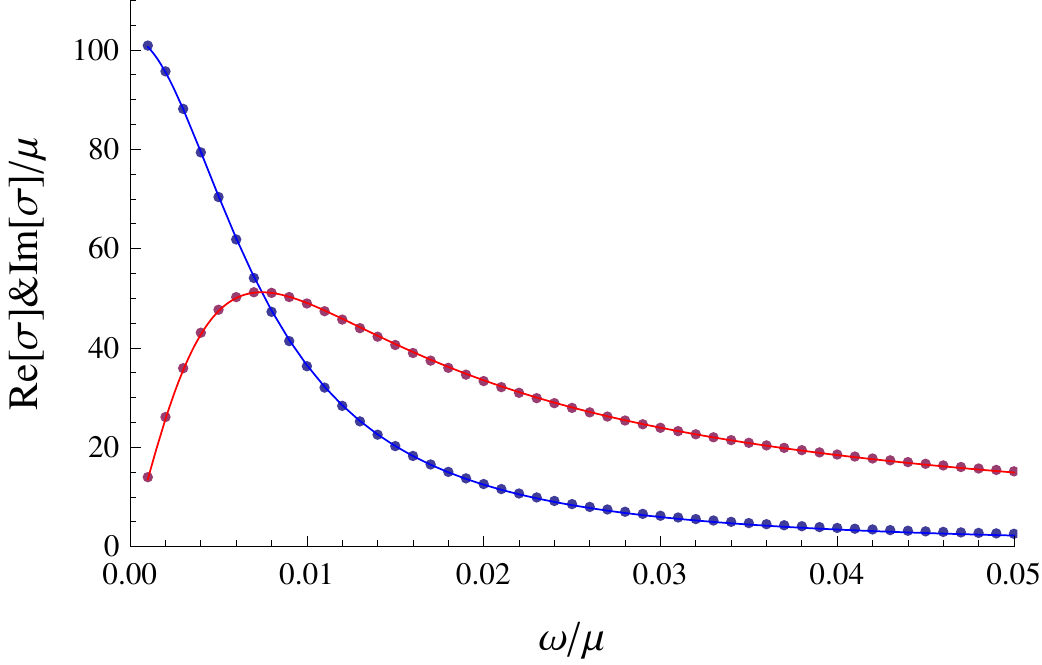}\hspace{0.5cm}
\includegraphics[scale=0.45]{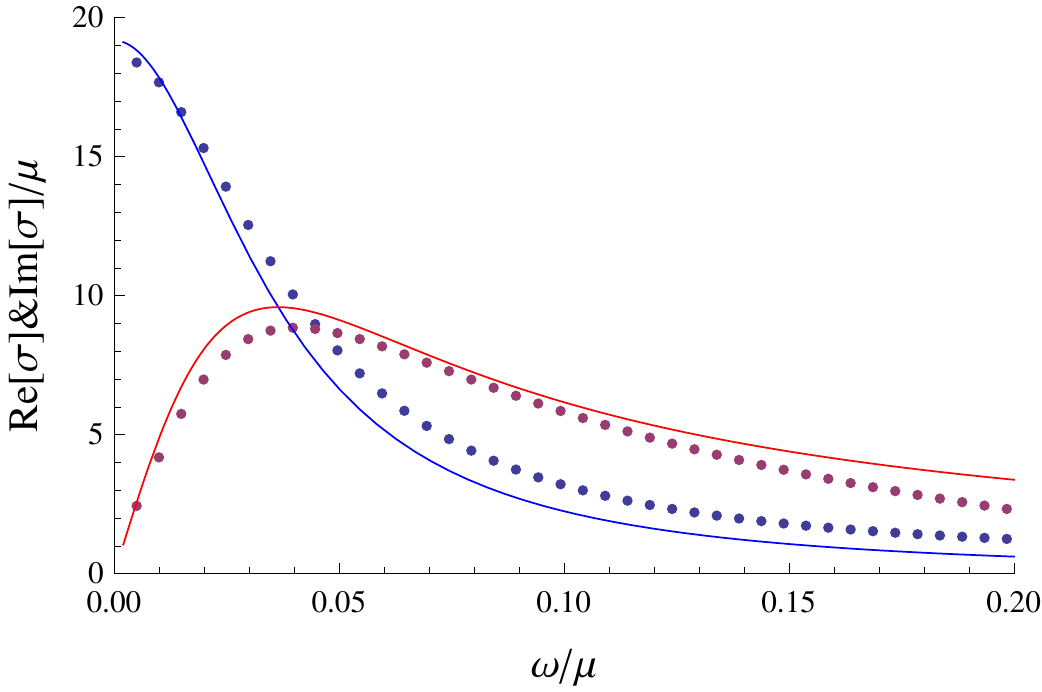}\hspace{0.5cm}
\includegraphics[scale=0.45]{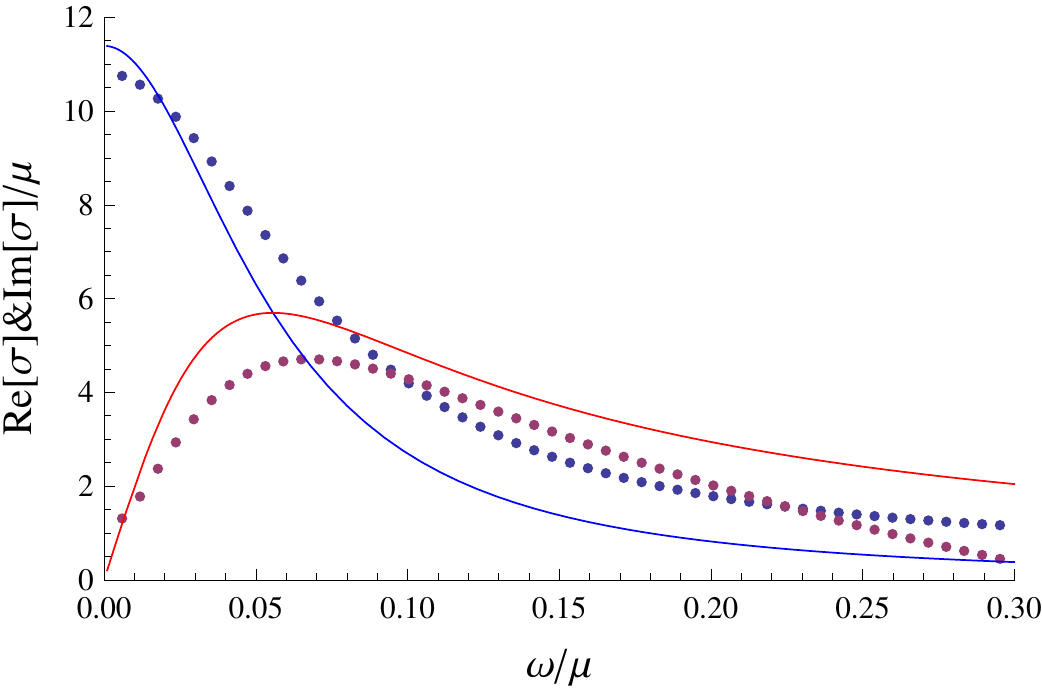}\hspace{0.5cm}
\includegraphics[scale=0.45]{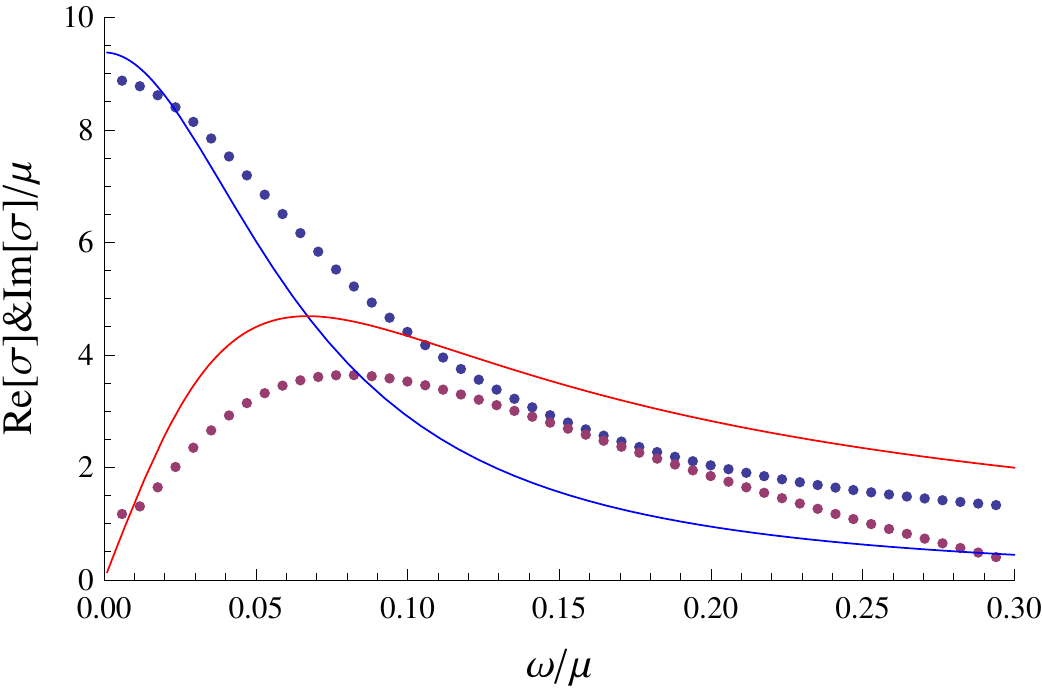}\hspace{0.5cm}
\includegraphics[scale=0.45]{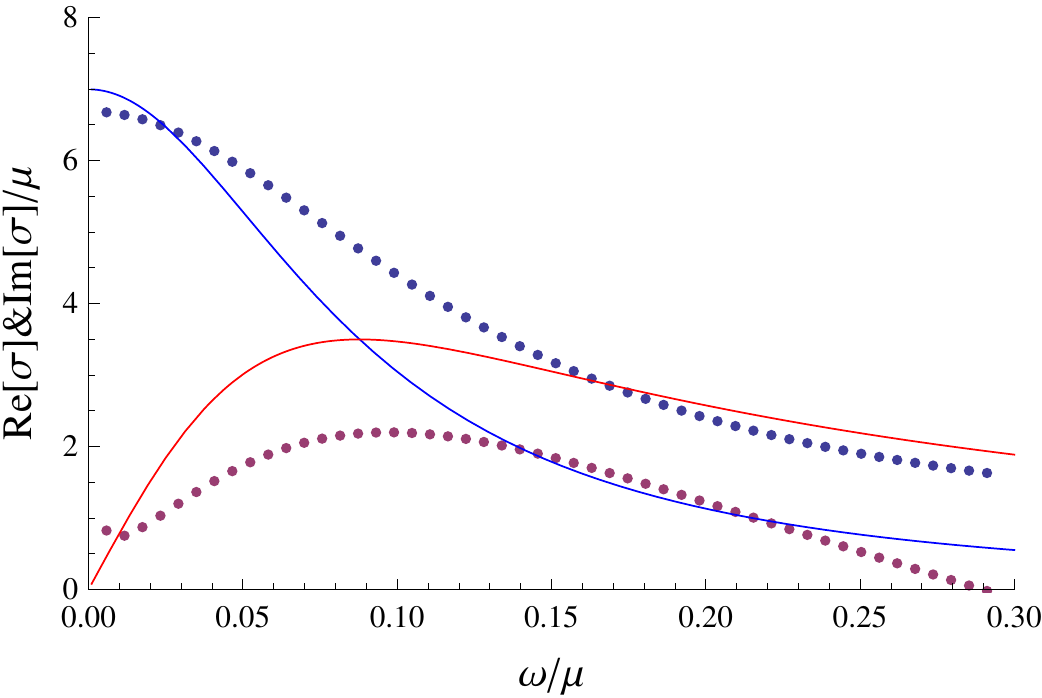}\hspace{0.5cm}
\caption{\label{solu}The optical conductivity as a function of
frequency for different wave numbers $a/\mu=0.1,0.2,0.5,0.7,0.8, 1.0$ from left
to right. The temperature is fixed at $T/\mu=0.2658$. The dots are
data, and the solid lines are fit to the Drude formula.  } }
\end{figure}


The optical conductivity is given by
\be
\sigma(\omega)=-\frac{i G_{Jz Jz}}{\omega}.
\ee
The numerical computation shows that the optical conductivity takes the form of the  Drude conductivity as
\be
\sigma(\omega) =\sigma_{Q}+ \frac{K \tau}{1-i\omega \tau},
\ee with $\sigma_{Q}=\rh e^{\frac{\phi(\rh)}{4}}$,  constant $K$, and relaxation time $\tau$.
Figure  \ref{solu} shows how the optical conductivity changes as the anisotropic parameter $a$ (i.e. the dissipation strength) changes. Since the anisotropic parameter $a$ is nonvanishing, the $1/\omega$ pole in the imaginary part disappears. As shown in Table \ref{Tablev3}, when $a$ becomes bigger, the maximum value of  the peak in the real part decreases, which is in good agreement with the $dc$ conductivity (\ref{DCsig}).   We also emphasize that for the case $a/\mu=0.5$, the numerical data deviate from the standard Drude model, implying that there is a coherent/incoherent transition \cite{kim}. As $a/\mu$ increases, significant deviations from the Drude model can be observed and the conductivity looks incoherent without a Drude peak.  We expect the ac thermal and thermoelectric conductivities with momentum relaxation also to show a Drude peak at small $a/\mu$ as given in Ref. \cite{kim}, and we  postpone such computation to a future study.

\subsection{DC Thermoelectric conductivities }
In this section, we provide an alternative calculation of the  dc electric conductivity, thermoelectric conductivities (Seebeck coefficients), and the thermal conductivity by using the method developed by Donos and Gauntlett \cite{donos}.
In this setup, we consider a slightly different form of the black bole fluctuations,
\be
\delta A_z=-E t+a_z(r),~~~
h_{tz}=h_{tz}(r),~~~
h_{rz}=h_{rz}(r),~~~
\chi=a ~z+\delta \chi_1(r),
\ee
where the temporal component of the four-potential $a_\mu$ corresponds to a constant electric field $E$ along the $z$ direction. The  equations of motion for these linearized fluctuations are given by
\bea
&& h''_{tz}-\bigg(\frac{g'_{tt}}{2 g_{tt}}+\frac{g'_{rr}}{2 g_{rr}}+\frac{g'_{zz}}{2 g_{zz}}-\frac{g'_{xx}}{ g_{xx}}\bigg)h'_{tz}+\bigg(\frac{g'_{rr}g'_{zz}}{2 g_{rr}g_{zz}}+\frac{g'_{tt}g'_{zz}}{2 g_{tt}g_{zz}}-\frac{g''_{zz}}{g_{zz}}+\frac{g'^2_{zz}}{2 g^2_{zz}}\nonumber\\[0.7mm]&&-\frac{g'_{xx}g'_{zz}}{ g_{xx}g_{zz}}-\frac{g_{rr}e^{2\phi}a^2}{g_{zz}}\bigg)h_{tz}+A'_t a'_z=0,\label{htz1}\\
&&a''_z+\bigg(\frac{g'_{tt}}{2 g_{tt}}-\frac{g'_{rr}}{2 g_{rr}}-\frac{g'_{zz}}{2 g_{zz}}+\frac{g'_{xx}}{ g_{xx}}\bigg)a'_z+\frac{A'_t}{g_{tt}}h'_{tz}-\frac{g'_{zz}A'_t}{g_{zz}g_{tt}}h_{tz}=0,\label{az1}\\
&&\frac{2 A'_t E g_{rr}}{g_{tt}}-2g_{rr}e^{2\phi}a \delta\chi'_1+\bigg(\frac{g'_{rr}g'_{zz}}{ g_{rr}g_{zz}}-\frac{g'_{tt}g'_{zz}}{2 g_{tt}g_{zz}}-\frac{g''_{zz}}{g_{zz}}+\frac{g'^2_{zz}}{2 g^2_{zz}}+ \frac{2g''_{xx}}{g_{xx}}-\frac{g'_{rr}g'_{xx}}{ g_{rr}g_{xx}}\nonumber\\[0.7mm]&&-\frac{g'_{xx}g'_{zz}}{ g_{xx}g_{zz}}+\frac{g'_{tt}g'_{xx}}{ g_{tt}g_{xx}}\bigg)h_{rz}=0.\label{hrz}
\eea
Note that the derivative of the scalar potential is given by $A'_t=-Q \frac{(g_{rr}g_{tt})^{1/2}}{g_{xx}\sqrt{g_{zz}}}$. The equation (\ref{hrz}) can be solved easily:
\be
h_{rz}=\frac{EQ \sqrt{g_{rr}}}{a^2 e^{2\phi} g_{xx}\sqrt{g_{tt}g_{zz}}}+\frac{\delta\chi'_1}{a}.
\ee
In order to solve the equations of motion for $a_z$ and $h_{tz}$, we need to impose proper boundary conditions for the fluctuation fields at the event horizon $r=\rh$ and at the conformal boundary $r\rightarrow \infty$.
We first assume that $\delta \chi'$ is analytic at the event horizon and falls off fast at infinity. Regularity at the event horizon can be obtained by switching to Eddington-Finklestein coordinates:
\be
v=t-\frac{1}{4 \pi}\ln(r-\rh).
\ee
In this coordinate system, the gauge field is determined by the regularity as
\be \label{azh}
a_z=-\frac{E}{4\pi T}\ln(r-\rh)+\mathcal{O}(r-\rh).
\ee
From Eq. (\ref{htz1}), we know that the regularity at the event horizon requires
\be \label{htzh}
h_{tz}=\frac{EQ \sqrt{g_{zz}}}{a^2 g_{xx}}\bigg|_{r=\rh}+\mathcal{O}(r-\rh).
\ee
Near the boundary ($r\rightarrow \infty$), we have the falloff of $a_z\sim J^z r^{-2}$, where $J^z$ denotes the charge density current. As to $h_{tz}$, from equation (\ref{htz1}), we can see that there are two independent solutions, one of which behaves as $\sim c_1 r^2$ and the other as $\sim r^{-2}$. We require that there be no sources associated with thermal gradients, and thus the coefficient $c_1$ should be vanishing. We also demand that $\delta \chi'$ fall off fast enough so that it has no contribution to the boundary value of $h_{rz}$ .

We now turn to the computation of the dc conductivity. We can see that the conserved charge density current is indicated by the nonzero Maxwell equation (\ref{az1}), so we can define the current as
\be
J^z\equiv\sqrt{-g}f^{rz}, ~~~f^{rz}=\frac{a'_z h_{tz}}{g_{rr}g_{zz}g_{tt}}+\frac{A'_t}{g_{rr}g_{zz}}.
\ee
We emphasize that the  charge density current derived from the  Maxwell equation is not identical to    the radially conserved quantity given in (\ref{pi}). Since the current is radially conserved, $J^z$ can be evaluated both at the horizon and at the boundary. The dc electric conductivity along the $z$ direction is expressed as $\sigma_{DC}=J^z/E$. By using (\ref{azh}) and (\ref{htzh}), we finally obtain the dc electric conductivity:
\be
\sigma_{DC}=\frac{g_{xx}}{ \sqrt{g_{zz}}}\bigg|_{r=\rh}+\frac{Q^2 }{a^2 e^{2\phi} g_{xx}\sqrt{g_{zz}}}\bigg|_{r=\rh}=\rh e^{\phi(\rh)/4}+\frac{Q^2}{a^2\rh^3}e^{-3\phi(\rh)/4}.
\ee
This result exactly agrees with (\ref{sigdc}) obtained in the previous section.

The conserved heat current $\mathcal{Q}$ is defined through introducing a two-form associated with the Killing vector field $K=\partial_t$, and it is assumed as the following form \cite{donos}:
\be
\mathcal{Q}=2\sqrt{-g}\nabla^r K^{z}+A_t J^z=\sqrt{\frac{g_{tt}}{g_{rr}g_{zz}}}g_{xx}\bigg(-g^{tt}h_{tz}\partial_r g_{tt}+\partial_r h_{tz}\bigg)+A_t J^z.
\ee
Notice that the quantity $\mathcal{Q}$ is also radially conserved.
We can evaluate $\mathcal{Q}$ at the event horizon:
\be
\mathcal{Q}=\sqrt{\frac{g^2_{xx}}{{g_{rr}g_{tt}g_{zz}}}}h_{tz} \partial_r g_{tt}\bigg|_{r=\rh}.
\ee
The Seebeck coefficient can be  obtained at the event horizon $r=\rh$ by using the expression
\be
\alpha=s_{DC}=\frac{1}{T}\frac{\mathcal{Q}}{E}=\frac{4\pi Q}{a^2 e^{2\phi}}\bigg|_{r=\rh}.
\ee
In order to calculate the thermal conductivity, we need to consider perturbations with a source for the heat current. A consistent choice of the linearized fluctuations takes the form
\bea \label{fluct}
&&\delta A_z=-E t+\zeta A_t t+a_z(r),\\
&& h_{tz}=-\zeta t \sqrt{g_{tt}}/\sqrt{g_{rr}}+h_{tz}(r),\\
&& h_{rz}=h_{rz}(r),\\
&& \chi=a z+\delta\chi_1(r),
\eea
where $\zeta$ is a constant. According to the holographic dictionary, the coefficient $\zeta$ corresponds to the thermal gradient $-\nabla_z T/T$ \cite{hartnoll,herzog, musso}. The choice of the fluctuations ensures that all the time-dependent terms drop out of the conserved current $J^z$ and $\mathcal{Q}$. The equations of motion for $h_{tz}$ and $a_z$ remain the same as given in (\ref{htz1}) and (\ref{az1}).
We can solve the equation for $h_{rz}$ and obtain
\be
h_{rz}=-\frac{Q(\zeta A_t-E)\sqrt{g_{rr}}}{a^2 e^{2\phi}\sqrt{g_{tt}g^2_{xx}g_{zz}}}-\frac{g_{zz}(g^{-1}_{zz}\zeta\sqrt{g_{tt}/g_{rr}})'}{a^2 e^{2\phi}\sqrt{g_{tt}/g_{rr}}}+\frac{\delta\chi'_1}{a},
\ee
where $\delta\chi_1$ can be a constant at the event horizon. The regularity at the event horizon requires the gauge field to take the form $a_z=-\frac{E}{4\pi T} \ln(r-\rh)+\mathcal{O}(r-\rh)$. It was suggested in Ref. \cite{donos} that the horizon regularity condition for $h_{tz}$ can be obtained by switching to the Kruskal coordinates
\be
h_{tz}=g_{zz}\sqrt{\frac{g_{tt}}{g_{rr}}}h_{rz}\bigg|_{r=\rh}-\frac{\zeta \sqrt{g_{tt}}}{4\pi T \sqrt{g_{rr}}}\ln(r-\rh)+\mathcal{O}(r-\rh).
\ee
Again, we impose the boundary conditions at infinity: $a_z\sim J^z r^{-2}$ and $h_{tz}\sim r^{-2}$. We emphasize that under the choice of fluctuations given in (\ref{fluct}), the form of the conserved current does not change. The conserved currents evaluated at the event horizon are given by
\bea
&&J^z=\bigg[E\bigg(\frac{g_{xx}}{\sqrt{g_{zz}}}+\frac{Q^2}{a^2 e^{2\phi}g_{xx}\sqrt{g_{zz}}}\bigg)+\frac{\zeta Q g'_{tt}}{a^2 e^{2\phi} \sqrt{g_{rr}g_{tt}}}\bigg]\bigg|_{r=\rh},\\
&&\mathcal{Q}=\bigg[E\frac{ Q g'_{tt}}{a^2 e^{2\phi} \sqrt{g_{rr}g_{tt}}}+\zeta \frac{g'^2_{tt}g_{xx}\sqrt{g_{zz}}}{a^2 e^{2\phi}g_{rr}g_{tt} }\bigg]\bigg|_{r=\rh}.
\eea
We finally obtain the dc thermoelectric conductivities in the $z$ direction:
\bea
&&\sigma_{DC}=\frac{\partial}{\partial E}J^z=\bigg(\frac{g_{xx}}{ \sqrt{g_{zz}}}+\frac{Q^2 }{a^2 e^{2\phi} g_{xx}\sqrt{g_{zz}}}\bigg)\bigg|_{r=\rh},\\
&&\bar{\alpha}=\frac{1}{T}\frac{\partial}{\partial E}\mathcal{Q}=\frac{4\pi Q}{a^2 e^{2\phi}}\bigg|_{r=\rh},\\
&&\alpha=\frac{1}{T}\frac{\partial}{\partial \zeta}J^z=\frac{4\pi Q}{a^2 e^{2\phi}}\bigg|_{r=\rh},\\
&&\bar{\kappa}=\frac{1}{T}\frac{\partial}{\partial \zeta}\mathcal{Q}=\frac{4\pi s T}{a^2 e^{2\phi}}\bigg|_{r=\rh}.
\eea
Note that we have used the notation $2\kappa^2=1 $. One interesting point we need to clarify is that $a$ plays the role of both anisotropy and momentum relaxation source. Hence, a finite $a$ is necessary for our whole computation.
\footnote{ It is clear that if the anisotropic parameter $a^2<0$, the $\bar{\alpha}$, $\alpha$ and $\bar{\kappa}$ will take negative values and thus become unphysical because negative thermal conductivity  means that thermal current can flow from lower temperature objects to higher temperature objects spontaneously.   From this point of view, we can see that unstable quasinormal modes in the bulk may result in unphysical transports of the dual field theory. It seems that we should abandon the case $\tau_{rel}<0$ on the field theory side.}

We would like to introduce the thermal conductivity at zero electric current, which is the usual thermal conductivity that is more readily measurable
$\kappa=\bar{\kappa}-\alpha \bar{\alpha}T/\sigma_{DC} $, and hence
\be
\kappa=\frac{4\pi s T e^{-\phi}\rh^4}{Q^2+a^2 e^{\phi}\rh^4 }\bigg|_{r=\rh}.
\ee
In conventional metals, the Wiedemann-Franz law holds, and thus the Lorentz ratio is given by $L\equiv \kappa/(\sigma T)=\pi^2/3 \times k^2_{B}/e^2$. This reflects that for Fermi liquids the ability of the
 quasiparticles to transport heat is determined by their ability to transport charge so the  Lorenz ratio is a constant. It has been observed that the Wiedemann-Franz law does not hold in the high-temperature regime with linear temperature-dependent resistivity in heavy fermions \cite{tanatar}, signaling the appearance of strong interactions. For our holographic setup, we expect similar non-Fermi behaviors.   We find that  the Lorenz ratios are given by
\bea
\bar{L}\equiv \frac{\bar{\kappa}}{\sigma T}=\frac{s^2}{Q^2+a^2 \rh^4 e^{\phi} }\bigg|_{r=\rh},\\
{L}\equiv \frac{{\kappa}}{\sigma T}=\frac{s^2 a^2 e^{\phi}\rh^6 }{(Q^2+a^2 \rh^4 e^{\phi})^2 }\bigg|_{r=\rh}.
\eea
Deviations from the Wiedemann-Franz law can be observed from the above equations and also as $a\rightarrow 0$, $\bar{L}$ and $\kappa$ approach finite, while $L$ goes to zero and $\bar{\kappa}$ diverges.

\section{Shear viscosities and viscosity bound}
For the anisotropic fluid considered here, the viscosity tensor $\eta_{ijkl}$ yields two shear viscosities out of five independent components \cite{landau}.
In the $u$ coordinate, we work with the $h_{u\nu}=0$ and $A_u=0$ gauges and consider linearized fluctuations of the form
$h_{xy}=e^{-i\omega t+i k_z z}h_{xy}(u)$ for the transverse shear viscosity, and $h_{xz}=e^{-i\omega t+i k_y y}h_{xz}(u)$ for the longitudinal shear viscosity.

For the transverse tensor mode $h_{xy}$, the equation of motion is given by
\ba
0={h_y^x}''-\frac{3}{u}{h_y^x}'+\frac{1}{2}\frac{\ch'}{\ch}{h_y^x}'+\frac{\cf'}{\cf}{h_y^x}'
-\frac{3}{4}\phi'{h_y^x}'+\frac{\cb'}{2\cb}{h_y^x}'- \frac{k^2_z  {h_y^x}}{\cf\ch}+\frac{\omega^2 {h_y^x}}{\cf^2\cb}.
\ea
We introduce the following notation \cite{mamo}:
\ba
\mathcal{N}^{\mu \nu}=\frac{1}{2\kappa^2}g_{xx}\sqrt{-g}g^{\mu\mu}g^{\nu\nu},
\ea
so that the equation of motion for $h_{xy}$ can be written as
\ba
\partial_u(\cn^{uy}\partial_u h_y^x)-k^2_z \cn^{zy} h_y^x-\omega^2 \cn^{ty}h_y^x=0.
\ea
The Green function is
\be
G_{xy,~xy}=\frac{\cn^{uy}\partial_u h_y^x}{h_y^x}.
\ee
The shear viscosity is defined as
\be
\eta_{xy,~xy}=-\frac{G_{xy,xy}}{i\omega}.
\ee
The flow equation for the transverse viscosity  is given by
\be
\partial_u\eta_{xy,~xy}=i\omega(\frac{\eta^2}{\cn^{uy}}+\cn^{ty})+\frac{i}{\omega}\cn^{zy}k^2_z.
\ee
The transverse shear viscosity is easily obtained by demanding horizon regularity:
\be
\eta_{xy,~xy}=(-\cn^{ty}\mathcal{N}^{uy} )^{\frac{1}{2}}\bigg|_{u=u_H}=\frac{e^{-\frac{5\phi_H}{4}}}{2\kappa^2 \uh^3}=\frac{s}{4\pi}.
\ee
For both prolate and oblate anisotropy, the transverse shear viscosities are exactly  $\frac{s}{4\pi}$, and thus the viscosity bound is satisfied.
However, for the longitudinal tensor mode, we have the equation of motion for
$h_{zx}$:
\ba
0={h_z^x}''-\frac{3}{u}{h_z^x}'-\frac{1}{2}\frac{\ch'}{\ch}{h_z^x}'+\frac{\cf'}{\cf}{h_z^x}'
-\frac{3}{4}\phi'{h_z^x}'+\frac{\cb'}{2\cb}{h_z^x}'- \frac{k^2_y  {h_z^x}}{\cf}+\frac{\omega^2 {h_z^x}}{\cf^2\cb}.
\ea
We can recast it as
\be
\partial_u\eta_{xz,~xz}=i\omega(\frac{\eta^2}{\cn^{uz}}+\cn^{tz})+\frac{i}{\omega}\cn^{yz}k^2_y.
\ee
The horizon regularity requires
\be \label{shear}
\eta_{xz,~xz}=\frac{s}{4\pi \ch(\uh) }.
\ee
For prolate black brane solutions with  $\ch(\uh)>1$ , the shear viscosity to entropy density ratio $\frac{\eta_{xz,~xz}}{s}=\frac{1}{4\pi \ch(\uh)}<\frac{1}{4\pi }$ violates the KSS bound
\footnote{ Intriguingly,  for the oblate black brane solution with $a^2<0$ in which $\ch(\uh)<1$, the shear viscosity to entropy density ratio $\frac{\eta_{xz,~xz}}{s}=\frac{1}{4\pi \ch(\uh)}>\frac{1}{4\pi }$  satisfies the KSS bound. However, in this case, the oblate anisotropy results in causality violation and then such a solution is unphysical. } \cite{kss,mamo,critelli}. We notice that
the form of (\ref{shear}) exactly agrees with the result obtained by Rebhan and Steineder in Ref. \cite{rehban}, which is the first example of a shear viscosity falling below the KSS bound in Einstein gravity with fully known gauge-gravity correspondence. One point that should be pointed out is that in (\ref{shear}) the factor $\ch(\uh)$ receives contributions from the gauge fields. Moreover, the transverse shear viscosity $\eta_{xy,~xy}=\frac{s}{4\pi }$ reproduces the universal value for Einstein gravity with isotropic horizon geometry, also agreeing with Ref. \cite{rehban}.

Comparing the shear viscosity obtained here with that of Einstein-Gauss-Bonnet gravity \cite{gb,gb0,gb1,gb2,gb3,gb4,gb5,gbx,gb6,gb7}, we can find that the anisotropic parameter $a^2$ plays the same role as the Gauss-Bonnet (GB) coupling constant. When the (GB) coupling
constant takes a positive value, the viscosity bound is violated, but there is no KSS bound violation for the negative-valued GB coupling.

\subsection{Causality analysis}
In this subsection, we will show how an oblate anisotropy  leads to pathological boundary field theory by exploring the causality analysis.
We know that the anisotropy parameter $a$ acts as an isotropy-breaking external source that forces the system into an anisotropic equilibrium state.
The $\theta$ parameter is dual to the type-IIB axion $\chi$ with the form $\chi=az$ in which $a$ only plays the role of anisotropy and does not add new degrees of freedom to the SYM theory. On the dual quantum field theory side, oblate solutions with imaginary $a$ look like a nonunitary deformation and could result in a negative field coupling. In the following, we give
a clear explanation how this occurs.

For the transverse mode $h_{xy}(t,u,z)$, we obtain the equation of motion for $h_{xy}$
\bea
\partial_u(\cn^{uy}\partial_u h_y^x)-k^2_z \cn^{zy} h_y^x-\omega^2 \cn^{ty}h_y^x=0,
\eea
with the following notation:
\bea
\cn^{\mu \nu}=\frac{1}{2\kappa^2}g_{xx}\sqrt{-g}g^{\mu\mu}g^{\nu\nu}.
\eea
To see the causality on the boundary, we simply assume $h^x_y=e^{-i\omega t+i k_z z+i k_{u}u}$. In the large-momentum limit, the effective geodesic equation can be recast as
$k^{\mu}k^{\nu}g^{\rm eff}_{\mu\nu}=0$.  The effective metric can be given by
\be
ds^2_{\rm eff}=\cf\cb(-dt^2+\frac{\ch}{\cf\cb}dz^2)+\frac{1}{\cf}du^2.
\ee
The local speed of light is given by
\be
c^2_g=\frac{\cf\cb}{\ch}.
\ee
In the standard Fefferman-Graham (FG) coordinate, the expansion of the functions $\cf$, $\cb$, and $\ch$ can be written as  \cite{mateos1,cgs1}
\bea
\cf &=&1+\frac{11 a^2}{24}v^2+\bigg(\cf_{4}+\frac{11a^4}{144}\bigg)v^4+\frac{7}{12}a^4 v^4 \log v+\mathcal{O}(v^6),\\
\cb&=&1-\frac{11 a^2}{24}v^2+\bigg(\cb_{4}-\frac{11a^4}{144}\bigg)v^4-\frac{7}{12}a^4 v^4 \log v+\mathcal{O}(v^6),\\
\ch&=&1+\frac{a^2}{4}v^2-\bigg(\frac{2}{7}\cb_4-\frac{173 a^4}{4032}\bigg)v^4+\frac{a^4}{6}  v^4 \log v+\mathcal{O}(v^6).
\eea
We can expand the local speed of light $c^2_g$ near the boundary $v\rightarrow 0$:
\be
c^2_g-1=-\frac{a^2}{4}v^2+\mathcal{O}(v^4).
\ee
As the local speed of gravitons should be smaller than 1 (the local speed of the boundary CFT), we require
\be
c^2_g-1=-\frac{a^2}{4}v^2\leq 0.
\ee
The above ansatz leads to $a^2\geq 0$.
Following the procedure of Ref. \cite{gb}, one can find that the group velocity of the graviton is given by
\be
v_g=\frac{ \bigtriangleup z}{ \bigtriangleup t}\sim c_g.
\ee
In a word, as near the boundary $c_g$ becomes greater than $1$, the propagation of signals in the boundary
theory with speed $\frac{\vartriangle z}{ \vartriangle t}$ could become superluminal.
Therefore, causality structure of this theory requires $a^2\geq 0.$\footnote{ We also note that for the longitudinal modes $h_{xz}(t,u,y)$, the local speed of light is exactly $1$, i.e. $c^2_g=1$}

\section{ Gubser-Mitra conjecture and ``Wall of stability"}
The Gubser-Mitra (GM) conjecture claims that gravitational backgrounds with a translationally invariant horizon develop a dynamical instability precisely whenever the specific heat of the black brane geometry becomes negative  \cite{GM}. The GM conjecture was later refined as working provided  that there is a unique background with a spatially uniform horizon and specified conserved charges  \cite{FGM}.
A holographic realization of the GM conjecture was given in Ref. \cite{buchel} by demonstrating that a tachyonic mode of the GM instability is dual to an imaginary sound wave in the gauge theory.

On the other hand, the wall of stability  refers to the regime $\tau_{rel}\geq 0$. This means that the total momentum density $T^{zt}$ of the field theory becomes unstable and grows in time when $\tau_{rel}<0$, because it absorbs momentum, rather than dissipating it.
Thus, the fluctuations will grow exponentially in time.  The wall of stability in fact imposes some constraints on the anisotropic parameter  from the dynamical side.

 It is our purpose in this section to consider the GM conjecture
by comparing the  dynamical  to the thermodynamic instabilities in our anisotropic system. For completeness of our study, we also extend our discussions to the massive gravity theory  \cite{vegh} and the Einstein-Maxwell linear scalar theory \cite{withers}.

\subsection{Dynamical and thermodynamic instabilities in the anisotropic background}
In Sec. III, we proved that the relaxation time $\tau_{rel}$ is proportional to $a^2$. That is to say, $\tau_{rel}$ is positive for the prolate anisotropy, as we already knew that
the prolate solution has a thermodynamic instability at smaller horizon radii. This means that the dynamical instability uncovered in this anisotropic
background is not correlated with the thermodynamic instability.

It would be interesting to examine the GM conjecture by considering the sound modes in our anisotropic media: whenever the specific heat of the prolate black brane is negative, the speed of sound in such a system should be imaginary. The speed of sound determined from the equation of state is given by
\be
v^2_s=\frac{\partial P}{\partial \mathcal{E}}.
\ee
The  thermodynamic potential  in the grand   canonical ensemble is found to be $G=-P_z= \mathcal{E}-Ts-\rho \mu-a \Phi$ \cite{cgs,cgs1}. The entropy density can be written as
\be
s=-\bigg(\frac{\partial G}{\partial T}\bigg)_{\mu,\Phi}=\bigg(\frac{\partial P_z}{\partial T}\bigg)_{\mu,\Phi}.
\ee
The specific heat can be defined as
\be
c_{\mu,\Phi}=\bigg(\frac{\partial \mathcal{E}}{\partial T}\bigg)_{\mu,\Phi}.
\ee
The speed of sound then can be expressed as
 \be v_s^2=\frac{s}{c_{\mu,\Phi}}.\ee
This implies that for the case $c_{\mu,\Phi}<0$, the speed of sound is purely imaginary, since the entropy density is always positive.
 It is clear that imaginary speed of sound  in the gauge theory is unphysical, and this unphysical  quantity is related to the tachyonic mode of the GM instability\cite{buchel}. This is the meaning of the holographic interpretation of the   GM conjecture presented in Ref. \cite{buchel}. Similarly, in our paper,
a negative relaxation time scale leads to  tachyonic modes in the bulk gravitational theory and unphysical thermal conductivity on the dual field side.

 We can in turn consider the case in which the relaxation time $\tau_{rel}> 0$ with prolate anisotropy where there are no dynamical instabilities for the shear modes. The transport coefficients are all defined as positive . However, at the smaller-horizon branch, the specific heat of  the black brane becomes negative. Although we did not find dynamical instabilities of the shear modes for such negative specific heat, we emphasize that this cannot preclude that there is  a dynamical instability in sound modes. Considering the complexity and difficulty in the computation of the sound modes, we defer the study of the sound modes in our setup and its relation with the GM conjecture to a future publication.

\subsection{Instabilities of the black brane in the massive gravity model }
The application of massive gravity in holography with broken diffeomorphism invariance in the bulk introduces a mass term for the graviton in such a way that one has momentum relaxation
in the boundary dual field theory. The action of the four-dimensional massive gravity model is given by  \cite{vegh,blake1,amoretti}
\bea
\label{massivelag}
S =& &\int d^4x\  \sqrt{-g} \left[ \frac{1}{2 \kappa_4^2}\left(R+\frac{6}{L^2}+\beta \left([\mathcal{K}]^2
-[\mathcal{K}^2]\right) \right)-\frac{1}{4 }F_{\mu \nu}F^{\mu \nu} \right]\nonumber\\
&+&\frac{1}{2 \kappa_4^2} \int_{z=z_{UV}} d^3 x\ \sqrt{-g_b} \  2 K \ ,
\eea
where $\beta$ is an arbitrary parameter having the dimension of mass squared and $(\mathcal{K}^2)^\mu_{\nu}\equiv g^{\mu\rho}f_{\rho\nu}$, $f_{\mu\nu}={\rm diag}(0,0,1,1)$.
The black brane solution in this massive gravity is given by
\bea
&&ds^2=\frac{L^2}{u^2}\bigg[-f(u)dt^2+dx^2+dy^2+\frac{1}{f(u)}du^2\bigg],\\
&&A_t=\mu(1-\frac{u}{\uh}),\\
&&f(u)=\frac{\gamma^2 \mu^2 u^4}{2L^2\uh^2}-\frac{\gamma^2\mu^2 u^3}{2L^2\uh}-\frac{u^3}{\uh^3}-\frac{\beta u^3}{\uh}+\beta u^2+1.
\eea
The black hole temperature is written as
\be
T=\frac{3}{4\pi \uh}-\frac{\gamma^2 \mu^2 \uh}{8\pi L^2}+\frac{\beta \uh}{4\pi}.
\ee
It is easy to check that for the case $\beta<0$, the local stability condition ${\partial T}/{\partial \uh}<0$ is always satisfied. This is to say, as the horizon radius $r_H=1/\uh$ increases, the black hole temperature goes up. However, for the case $\beta>0$, there is a branch of black brane solutions having ${\partial T}/{\partial \uh}>0$. That is what we mean, the instability of the black brane because the heat capacity could become negative.
The heat capacity is computed in the usual way:
\be
c_{\rho}=\frac{\partial\mathcal{E}}{\partial T}=\bigg(\frac{\partial \mathcal{E}/\partial \uh}{\partial T/\partial \uh}\bigg)_{\rho}=\left(-\frac{3 L^2}{\uh^4 \kappa^2_4}-\frac{\beta L^2}{\uh^2 \kappa^2_4}-\frac{\mu^2}{2 \uh^2}\right)\frac{1}{\frac{\beta}{4\pi}-\frac{\kappa^2_4\mu^2}{8L^2 \pi}-\frac{3}{4\pi\uh^2}}.
\ee
Note that if ${\partial T}/{\partial \uh}>0$, the heat capacity becomes negative, since $\partial \mathcal{E}/\partial \uh<0$ for the entire range of the parameters (see also Refs. \cite{chp,lu1,lu2,lu3}). We find that for $\beta>\frac{6 L^2+ \kappa^2_4\mu^2 \uh^2}{2L^2 \uh^2}:=\beta_{c}$, the black brane is thermodynamically unstable.

It is interesting to note that the thermodynamic instability uncovered here is related to the dynamical instability of the dual fluid.\footnote{We would like to thank Richard Davison for figuring out this point.}
The momentum dissipation rate determined in terms of the graviton mass and the equilibrium thermodynamical quantities is given by \cite{davison}
\be
\tau^{-1}_{rel}=-\frac{s\beta}{2\pi(\mathcal{E}+P)}.
\ee
The case $\beta>0$ corresponding to $\tau_{rel}<0$ means that the fluid turns to gaining momentum. The amplitude of the shear diffusion mode thus will  grow exponentially in time and leads to instability of the system.
Therefore, the wall of stability imposes more tight constraints on the parameter $\beta$ than the black brane thermodynamics, and the regime of the dynamical instability does not  coincide with the regime of the thermodynamic instability completely. This implies that our result  provides a counterexample to the  GM conjecture:  the dynamical instability occurs even when the black brane is thermodynamically stable, as we  saw above in  the window $0<\beta<\beta_{c}$.

\subsection{Instabilities of the black brane in the Einstein-Maxwell linear scalars theory}
In the holographic model consisting of Einstein-Maxwell theory with linear scalar fields, momentum relaxation can be realized through spatially dependent sources for operators dual to neutral scalars.  The five-dimensional action can simply be written as
\begin{equation}
	S = \int_M \sqrt{-g} \left[ R - 2 \Lambda - \frac{1}{2} \sum_{I}^{4-1} (\partial \chi_I)^2  - \frac{1}{4} F^2 \right ] d^{4+1} x
	- 2 \int_{\partial M} \sqrt{-\gamma} K d^4 x, \label{the model}
\end{equation}
where $\Lambda=-12/(2l^2)$ and $\chi_I$ denotes an axion field. The resulting black branes are homogeneous and isotropic:
\bea
&&ds^2=-f(r)dt^2+\frac{dr^2}{f(r)}+r^2 dx^idx^i,~~~ A_t=\mu(1-\frac{r^2_H}{r^2}),~~~\chi_I=\beta_{Ii}x^i,\\
&&f(r)=r^2- \frac{\beta^2}{4}-\frac{m_0}{r^2}+\frac{\mu^2 r^4_{H}}{3 r^4},~~~ m_0=r^4_H(1+\frac{\mu^2}{3 r^2_H}- \frac{\beta^2}{4 r^2_H}).
\eea
The temperature of the black brane is given by
\be
T=\frac{1}{4\pi}\bigg(4 r_H -\frac{\beta^2}{2r_H}-\frac{2\mu^2}{3r_H}\bigg).
\ee
One may notice that through an analytical continuation $\beta \rightarrow i \beta$, the black brane temperature here becomes exactly that of the massive gravity.
We stress that the sign of the $\beta^2$ is arbitrary. The specific heat  can be read as
\be
c_{\rho}=8\pi \rh^3 \left(4-\frac{\beta^2}{2 \rh^2}+\frac{2\mu^2}{3 \rh^2}\right)\frac{1}{4+\frac{\beta^2}{2 \rh^2}+\frac{2\mu^2}{3 \rh^2}}.
\ee
It is easy to find that  for  any given temperature $T>0$ and positive $\beta^2>0$, the heat capacity is positive. If the constant $\beta$ is analytically continued to an imaginary value, the black brane solution will
generate an unstable branch with $c_{\rho}<0$.  Also, the  momentum dissipation rate is given by $\tau^{-1}_{rel}=\frac{s\beta^2}{2\pi(\mathcal{E}+P)},$ with a sign difference with that of massive gravity. Therefore, considering the parameter $\beta^2$ in the range $ \beta^2  \in (- \infty,\infty) $, we can conclude that the regimes of thermodynamic and dynamical instabilities do not equal each other.

\section{Conclusions}
In summary, we have investigated various aspects of the dynamics of the linear perturbations, in particular the effect of the relaxation of momentum upon various observables,  in the spatially anisotropic $\mathcal{N}=4$ super-Yang-Mills theory dual to the action (\ref{5action}). We computed the dc thermoelectric conductivities analytically. The optical conductivity was obtained through numerical computation.
  Actually, we uncovered a very interesting mechanism: the underlying model is anisotropic, and coherent/incoherent metal transition is realized in our model. This is because  we only calculated metric perturbations with $t$ and $z$ components of the gauge field; the optical conductivity was obtained along the  $z$ direction which is proven to be momentum dissipated. But in the $x$ and $y$ directions, it is still a metallic phase (see also Ref. \cite{donnos1}).   We also computed the shear viscosities and checked the viscosity bound for the prolate anisotropy.
   Finally, we examined the relations between the GM conjecture and
the wall of stability by comparing conditions for dynamical instabilities with  conditions for thermodynamic instabilities in massive gravity and Einstein-Maxwell linear scalar theory. It is still an open question to us whether the GM conjecture is strictly obeyed by this anisotropic system. It was noticed in Ref. \cite{liwei} that anisotropically deformed $\mathcal{N}=4$ Yang-Mills plasma at zero chemical potential has low-temperature instabilities, leading to a new ground state with anisotropic scaling  \cite{banks}. It would be interesting to find a new black hole solution at nonzero chemical potential by considering the influence of such an instability and to study the transport properties, because this instability may relate to the GM conjecture discussed here.

The optical conductivity matches  the Drude model for small $a/\mu$ and exhibits incoherent behavior at significantly higher values of $a/\mu$.
  In Ref. \cite{fang}, it was found that the probe fermions in this anisotropic background are a non-Fermi liquid type without well-defined quasiparticle sates. Nevertheless, our results in this paper  show that the dual system
  shows Drude-type behavior with small anisotropy in ac conductivity, although there are no
  coherent quasiparticle states.
 In the small-black-hole-radius branch,  which is unstable, the dc electric conductivity shows the strange metal behavior with a linear temperature resistivity. It is a future project to deform the present model to stabilize this branch,  which  is very motivating phenomenologically.

 The ratio of the shear viscosity to entropy density violated the viscosity bound for the prolate black brane solution. On the other hand, for the momentum relaxation case, the oblate black brane solution is useless in calculating the conductivities: It may give unphysical, negative thermoelectric conductivity.
   In the future, it might be interesting to consider the holographic transports of anisotropic black branes with higher-derivative gravity terms by adding chemical potential to the model constructed in Ref. \cite{tran} and investigate the diffusive bound as in Ref. \cite{Amoretti3}.

\section*{Acknowledgements}
We thank A. Buchel, R. G. Cai, R. Davison, B. Gouteraux, B. S. Kim,  J. X. Lu,  N. Iqbal, S. F. Wu, and
especially K.-Y. Kim for the useful discussions.
X. -H. G. was partially supported by NSFC,
China (Grant No.11375110). Y. L.  was partially supported by NSFC,
China (Grant No.11275208). S. -J. S. was supported partially by the NRF, Korea (Grant No.NRF-2013R1A2A2A05004846).
Y. L. also acknowledges the support from the Jiangxi
Young Scientists (JingGang Star) program and the 555 talent project of Jiangxi Province.

\end{document}